# Improving Usability of User Centric Decision Making of Multi-Attribute Products on E-commerce Websites

**Roquia Mushtaq[1], Naveed Ahmad[2], Aimal Rextin[3], Muhammad Muddassir Malik[4]**

[1]Department of Computer Science, Air University, Islamabad, Pakistan
*(roquia.mushtaq@mail.au.edu.pk)*
[2]Department of Computer Science, National University of Computer and Emerging Sciences, Islamabad, Pakistan
*(naveed.ahmad@nu.edu.pk)*
[3]Department of Computer Science, COMSATS University, Islamabad, Pakistan
*(aimal.rextin@comsats.edu.pk)*
[4]School of Electrical Engineering and Computer Science, National University of Science and Technology, Islamabad, Pakistan
*(muddassir.malik@seecs.edu.pk)*

**Abstract** The high number of products available makes it difficult for a user to find the most suitable products according to their needs. This problem is especially exacerbated when the user is trying to optimize multiple attributes during product selection, e.g. memory size and camera resolution requirements in case of smartphones. Previous studies have shown that such users search extensively to find a product that best meets their needs. In this paper, we propose an interface that will help users in selecting a multi-attribute product through a series of visualizations. This interface is especially targeted for users that desire to purchase the best possible product according to some criteria. The interface works by allowing the user to progressively shortlist products and ultimately select the most appropriate product from a very small consideration set. We evaluated our proposed interface by conducting a controlled experiment that empirically measures the efficiency, effectiveness and satisfaction of our visualization based interface and a typical e-commerce interface. The results showed that our proposed interface allowed the user to find a desired product quickly and correctly, moreover, the subjective opinion of the users also favored our proposed interface.

## Introduction

According to the global report on e-commerce, it is estimated that the current growth rate of e-commerce market has increased up to 24.8 percent globally [1]. In addition, by the year 2021 it is expected that a total of 2.14 billion people will buy products and services online. It can be safely said that e-commerce websites are an emergent trend of today [1][2]. Particularly, in e-commerce settings, a user searches for a type of product say phone or camera or service like holiday package. The user is returned numerous products and each product has multiple *attributes* such as brands, price, rating, and various specification etc. [3][4]. The process of product purchase decision making involves identifying the objective, collecting information, and assessing alternative options [5][6]. However, it is not easy for users to determine the suitable product service due to high information load [4][7].

Product/services data is present in databases and is generally in ordinal or quantitative form and hence suitable for computer-based processing. An overview of the literature reveals two diverging approaches to help users find the best product or service. The first approach is algorithmic techniques to shortlists products/services based on a user's demographic information, customers, past searching and buying behavior [8][9][10]. The second approach is the visualization-based approach that shows the data of interest to the user in pictorial form. We will discuss this approach in a slight detail below.

Information Visualization (IV) is defined as *"the use of computer-supported, interactive, a visual representation of data to*

*amplify cognition*", whereas the term cognition refers to the ability to perceive data by a human mind [11]. Huang *et al.*[12] argued that visually representing data makes it possible to perform some tasks with simple perceptual operations, they designed a human cognitive to process visual information more quickly and effectively [12]. Moreover, visualization can also function as an "external memory" of a human mind [13], thus helps in reducing cognitive load on users. However, despite its promise, it seems that it is used less frequently in e-commerce. One can also find a few examples of its use by commercial ecommerce websites like cylindrical bar visualization by eBay.com [14] and visualization based flight reservation by hipmunk.com. One can also find many studies to help users in specific domains like social networks [15], flight reservation systems [16], music video systems [10] etc. However, these techniques are difficult to generalize for a general-purpose ecommerce decision making and even latest literature suggests the use of elementary visualization techniques like parallel coordinates, scatter plot and tabular visualization in an ecommerce setting [9].

We prefer the visualization based approach as various reports [17] have shown that sometimes relying too much on algorithms can have undesired effects. One example of undesirable consequences of relying on algorithm is the fact that Amazon recently discontinued an AI-based CV shortlisting system when they discovered that it was biased against female applicants [18]. However, our research is based on the premise that control of finding products and services should be given to a human rather than to an algorithm. The difference between the two approaches can be seen in Figure 1. In the Algorithmic approach the final output to the user will be a relatively short list of products based on his/her previous shopping history. While the second part shows the Visualization Approach, in which the user is presented with a large subset of the data which he and she can interact with to discover the desired information. In other words, we recall here that presenting a small subset of the data might result in the exclusion of a relevant product, so instead we present a large subset of data in visual form. This allows the allowing users to perform simpler perceptual operations [12] and reducing the cognitive load on a limited human memory [13][19] whereas in algorithmic approaches the output is controlled by the rules of the algorithm.

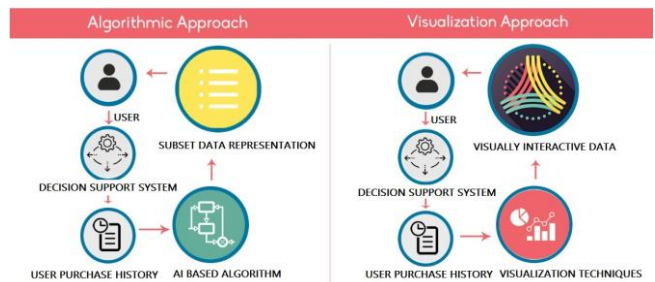

*Figure 1: Difference between the two approaches. We can see here that in the algorithmic approach an algorithm suggests to the user which products are likely to interest the user. While in the in the visualization approach a large part of the data is presented to the user and the decision making is done by the user spontaneously.*

Latest research recommends generating a picture of all available data in a single-go using elementary visualization techniques like scatter plot or parallel coordinates [9]. It might not be easy for a user to process such a visualization as there can be hundreds of products each with numerous data points as a result of any search query. Moreover, as we will see in the next section, a user generally decides on a product by gradually shortlisting the product until a decision is reached. Hence, we have the following two research questions:

1. **RQ1:** Can we propose visualization-based solution that help users find the most suitable product or services? We note that the product/services can be of different type of smartphones, computers, camera etc. We assume that each product or service has various associated attributes and at one time the user needs to compare products from a single type. We would also like that this visualization closely mimic the natural search and shortlisting that is done by a user.

2. **RQ2:** Quantify the benefit that an user will obtain from such a visualization-based interface [20].

The problem is to make the task of finding the optimum multi-attribute product/service easier for the user on a typical e-commerce website, hence, enhancing its usability. Here *multi-attribute* product or services means having various attributes of available products [19]. It is also important to note that the optimum product would differ from user to user as the desired product by definition is dependent on the various emotional, financial and other subjective reasons. Our proposed solution is a four step visualization pipe to help the user in finding the optimal product. These steps are: *filtering:* the user filters the product attributes for further consideration, *comparative view:* the user can perform comparison of the selected product attributes through a scatter plot visualization, *product view:* its objective is to provide quick detailed description of the product attributes for further consideration, *comparative visualization:* shows graphical comparison of selected product attributes to the user. Our proposed solution achieves its objective by helping the user understand the data with the help of pictorial representation of the data. Although, one can measure the reduction in cognitive load by gathering bio-medical measurements like EEG. Since our objective is to improve usability of an e-commerce interface, we adopted the method used in Human Computer Interaction by measuring the usability of the proposed visualization based interface as done in [21][22]. We consider three usability metrics [23][24]: *efficiency* is how quickly a user can complete a task, *effectiveness* is if the task was completed correctly or not; and *satisfaction* is how satisfied the user feels after using an interface.

The rest of the paper is organized into five sections: Section II presents the extensive literature review on de decision-making process. In Section III, we propose our 4-step visualization interface to help users in selecting an ideal product. Section IV describes the evaluation while section V discusses the results and analysis of the proposed visualization interface. Finally, Section VI concludes the paper and identifies avenues for future research.

## Related Work

As discussed in Section I, the number of online consumers is significantly increasing along with the products available on e-commerce website resulting in product selection dilemma i.e. finding the optimal product. In order to further understand this problem, this section presents literature on e-commerce decision making and existing solutions to facilitate this process.

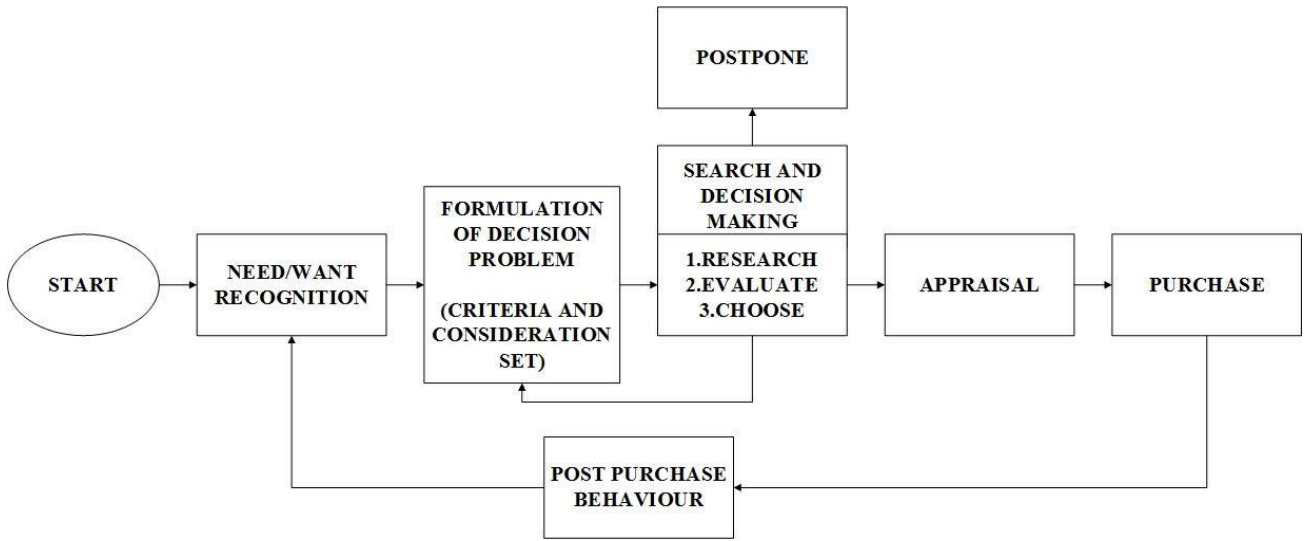

*Figure 2: Karimi's et al. framework for consumer purchase Decision-Making.*

### A. Electronic Consumer's Decision-Making Process

Consumer decision-making is defined as the "careful weighing and evaluation of utilization or functional product attributes to arrive at a satisfactory decision" [25]. Nicosia *et al.* [26] proposed a multi-stage linear model through which an online consumer decides to buy a product. These stages include: need recognition, searching, evaluating different alternatives available to him/her, purchase and post-purchase behavior [20][27]. It is observed that an online shopping portal can only help in the $2^{nd}$ and $3^{rd}$ step and the remaining steps are internal to the customer. The most critical step is consumers gradually formulating their expectations by comparing and reviewing attributes before the final purchase [26][28][29][30]. Studies indicate that online customer gradually shortlists a small set of products for detailed comparison, called the consideration set [28][19]. In order to construct a consideration-set, it is essential that an e-user overview all the relevant product attributes, such as their functionalities, prices etc. Similarly, Karimi *et al*. [20] proposed a new framework shown in Figure 2 that extends the linear decision-making process mentioned earlier. This model provides both a perspective and a descriptive view of the product decision-making process of an online consumer. This model proposes that a user decides to purchase by going through the following stages: need recognition, forming a consideration set, evaluation and appraisal. Hence, products are gradually shortlisted until he/she finds a satisfactory product. This short listing is done based on some criteria that is dependent on the user. Our proposed solution closely mimics the second and third stage of model proposed by Karimi *et al.* [20] as it allows the user to initially filter the desired product attributes and allowing them to make comparison of those selected attributes for further consideration before deciding the final product to purchase.

Karimi *et al.* [20] also proposes that there are two types of online customers: satisfiers and maximizers. Satisfiers look for a good enough product while maximizers try to find the best possible product. It has been seen that maximizers search extensively [31][29] and spend a lot of time evaluating various products to find the best possible product. Moreover, they consider all information presented to them and seriously consider a larger set of alternatives [30]. Karimi *et al*. [20] also showed empirically that maximizers take the most time before finalizing the product to be purchased.

Hence, we can conclude that maximizers will want to evaluate many options and then will gradually shortlist until a final decision is made. We note here that this decision making is closely related to the research area of *Multi-Criteria Decision Making (MCDM)* that aims to devise strategies to helps in decision making [32] including ordering or classifying alternatives [9]. However, the following are two key differences between MCDM and e-commerce decision making:

1. There can be finite or infinite alternatives and all of them might not be known in advance.

2. Users in ecommerce websites want to spontaneously decide, rather than following a strict procedure.

## B. Information Load in E-commerce

In terms of online purchasing, *information load* refers to "the amount of information conveyed through online product presentation that influences cognitive processing of a human mind" [33]. Online decision making has a cognitive element [3] as the total cognitive ability of a human mind is inversely proportional to overall information load [34]. Jacoby *et al.* [35] in their research studied relationship among the amount of information and the level of satisfaction when shopping on website. Their results indicated that a user has a lower ability to choose an ideal product when the amount of information is overwhelming. While purchasing online users not only suffer from product information overload such as product specifications, prices, reviews etc. but also deals with the multiple types of product representations such as visual or textual display.

Hong *et al.* [36] in their research examined two design features of displaying data on e-commerce websites. The presentation mode (textual vs visual mode) and the information format (array vs list format) while studying their impact on users. They concluded that image-based representation conveys more information to users than text and array-based representation. Similarly, Li *et al.* [37] in their research inferred from experimental results that visual display of product information yields better results than textual representation in enhancing users overall performance while reducing their cognitive load on shopping portals. Moreover, numerous studies investigated the ideal level of information that should be presented to the user for an optimal purchase decision [34][35].

Hence, a user who is a maximizer, i.e. looking for the best possible product must consider several attributes. For example, the user may be searching for smartphone with the best camera and screen size yet in under a certain price threshold. It is not easy for the maximizers to compare the various attributes [38][39] of hundreds of products and brands [36][37][40] simultaneously.

## C. Using Visualization to Resolve Information load

We recall that maximizers are e-customers who search extensively on e-commerce websites for the best possible product or service. This search is not easy given the large number of possible options available each with multiple attributes. We now wanted to search literature on how visualization can be used to make this search more efficient for the user. We first searched the published literature by snowballing and applying various search query. We couldn't find many papers that were directly related to our specific research problem. So, we applied a more formal strategy. We first chose list (https://sites.umiacs.umd.edu/elm/2016/01/21/infovis-venues/) of twelve journals and twenty conferences of Human Computer Interaction. We then went systematically through all papers published between 2015 and 2019 in the journals and conferences that were more relevant to our study i.e. Information Visualization in decision-making process. From which, we found 20 most relevant papers to look in greater depth. The summary of literature survey using both strategies is given below.

An overview of the literature reveals two diverging approaches to solve information overload problem. The first approach is to use an algorithm to suggest of products and services to the user. These results in an intelligent and efficient Decision Support Systems with Business intelligence tools embedded in them. An example of such systems is recommender systems are widely used in e-commerce portals [41][42]. These systems makes the overall process of product searching less time-consuming. However, such systems fall short when one wants to let the user decide by allowing the user to compare and shortlist to decide on the best product or service.

Research indicates that this limitation can probably be overcome through Information Visualization. Information visualization allows large datasets to be pictorially represented. Since human mind can quickly and easily process visual information as compared to textual data so it's easier to understand the overall trend in the data. We note here that on a computer the visualization is not always static but can be also changed by the user interacting with it. Furthermore, in order to be useful, [43] recommends that an interactive visualization should provide the following:

1. Provide an overview of the whole data,

2. Allows user to zoom or filter to focus on the information of interest,

3. And provides further details not available in the visualization when needed by the user.

Several studies in this area focus more on finding domains where visualization can make the task of user easier. This category is denoted as *application paper of visualization* by [44], these are papers that "examines and discusses the effectiveness of the visualization methods for a particular application". In this regard, visualization has been applied to vast number of domains, these include, among many others the following domains: social networks [15], flight reservation systems [16], music video systems [10], financial crime detection [45], tax evasion discovery [46], discovering required information in search results [47], exploring GitHub issues [47], information on infectious disease outbreak [48], decision making [49], finance [50][51], software release planning [52], health [52] and analysis of security risks [53] etc.

So, visualization makes it easier to compare and get products insight more clearly [19][54]. Chen *et al.* [19] in their research tried to utilize data visualization approach to improve the overall experience of online car buyers on product searching by allowing them to get an insight of their search results under diverse and multiple attributes. This searching interface is composed of parallel coordinate One can find many other similar examples of a wide variety of domain specific e-customer decision making applications such as *flight reservation systems* [16], *lightning design* [55], *house selection* [56], *film selection* [57] etc. However, these solutions are designed for a specific product or services e.g. cars and are not generalizable. Since a general e-commerce portal like Amazon or eBay has a variety of products, so we are interested in domain neutral visualization that helps a user's decision making.

General purpose visualization-based e-customer decision support systems generally use elementary visualization techniques like scatter plot matrix [58] ,bar charts [59][60] parallel coordinates [61] etc. In a very recent work, Dimara *et al.* [9] evaluated commonly used elementary visualization techniques for multi-attribute decision making. They found that all visualizations performed almost similarly in terms of their ability to allow a user to find the best product. However, Tabular Visualization, i.e. a color-coded table with text was the fastest. We couldn't not find a study that mimics the user decision making process, i.e. gradually shortlist products until a decision is made.

***Research Gap:*** We can see from the above that maximizers need to search extensively for the best possible product or service. However, there have been limited studies in this regard. In a very recent work [9] investigates how different elementary visualization perform for multi-attribute decision making. They found that tabular representation performs best in terms of time and that the effectiveness of all basic visualization was similar. However, these elementary visualizations still show a lot data points to the user. Thus, making it more time consuming for the user to decide upon the best product or service. Dimara *et al.* [9] mention that a visualization tool that combines various elementary visualization is probably needed. We also recall that an e-customer gradually shortlists products to a consideration set [28] and then the user makes the final decision. However, to the best of our knowledge such a *visualization pipeline* has not yet been proposed. We note that such a pipeline would work for almost all kinds of products and services whose majority of attributes are either ordinal or quantitative in nature.

## Proposed Interface

To deal with the e-commerce consumer's related shopping issues discussed above, we propose a visualization pipeline for such websites. Our workflow aims to streamline the search process through visualization, where product information containing multiple attributes can be processed interactively to shortlist, compare and choose a product. Here an *attribute* of a product is a characteristic that defines a product and can affect a consumer's purchase decision for example in case of smartphones; the attributes can be RAM, camera, price etc. Our proposed solution is designed with an objective to improve the overall process of product decision making for users by representing the products attributes in a visual format. This 4-step visualization pipeline helps the user gradually filter products until he/she has shortlisted a few products for detailed consideration. The user can then select the product he/she prefers for purchase. We note here that a user might like a comparison with the similar leading products or with leading brands. However, it is beyond the scope of this study as we focus more on how easy we can make it for the user to search a product with some specific criteria on various attributes of the products. Another important aspect to note is the operational target platform, the proposed visualization pipeline is not limited to any single platform and will work on all devices including PCs, laptops, touch screen devices etc. Figure 3 shows the four-step process on which our solution is based.

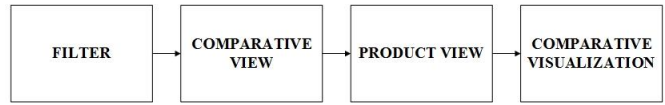

*Figure 3: Proposed visualization pipeline*

Each of these steps is explained in detail below.

1. *Filtering*

In this first step, we present product attributes in a compact form using the dynamic wheel visualization technique, as shown in Figure 4.

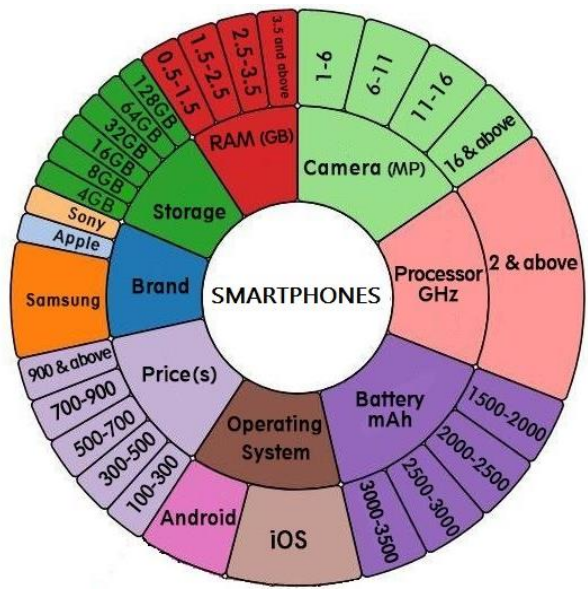

*a*

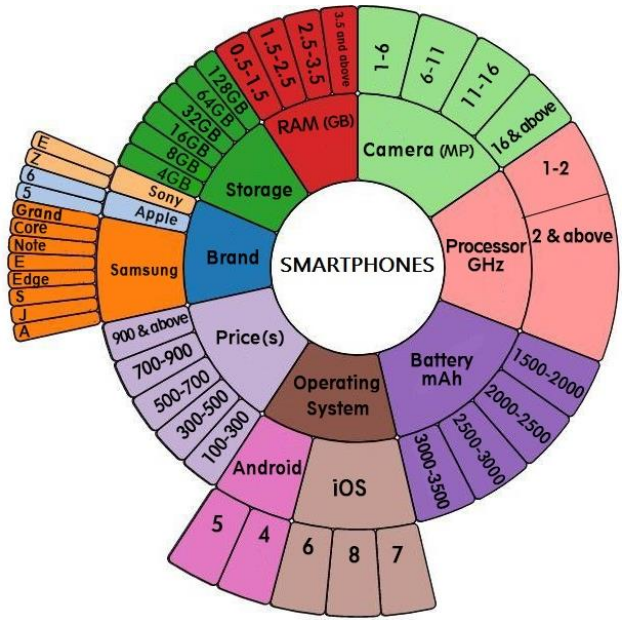

*b*

***Figure 4:** a) Product (Smartphone) attributes; b) Expanded attributes 'Operating System' and 'Brand'*

Figure 4 (a) shows the attributes and sub attributes of smartphones. This dynamic wheel visualization will change for every different product and automatically select attributes and sub attributes of the type of product user is looking for. A user can interact with this visualization in two ways. Firstly, user can expand the attributes further and can further explore any sub attribute, for example Figure 4 (b) shows expanded sub attributes in the brand Samsung, such as Grand, Edge etc. Secondly, user can easily shortlist the desired attributes by clicking on it. The objective of filtering step is to help users to select attributes that are important to him/her. For example, a user purchasing a smartphone online, might only consider camera resolution, screen size, storage capacity and is not concerned with other attributes. Such a user will only select the important attributes; hence this dynamic wheel visualization helps him/her to reduce the clutter of numerous unimportant attributes and let the user focus only on important attributes in future steps.

2. *Comparative View*

After filtering out the unnecessary smartphone attributes the user can further perform attribute comparison among selected ones. This filtering is reflected in a scatter plot visualization technique as shown in

Figure **5** where initially all the models are shown but as the user filters less and fewer gadgets are shown. The user can change the x and y axis of the scatter plot to the attributes of prime interest. Such visualization makes it very easy to identify which devices are outperforming others in user's attributes of interest [4]. The user can then right click on the desired smartphone to add it to the compare bucket list.

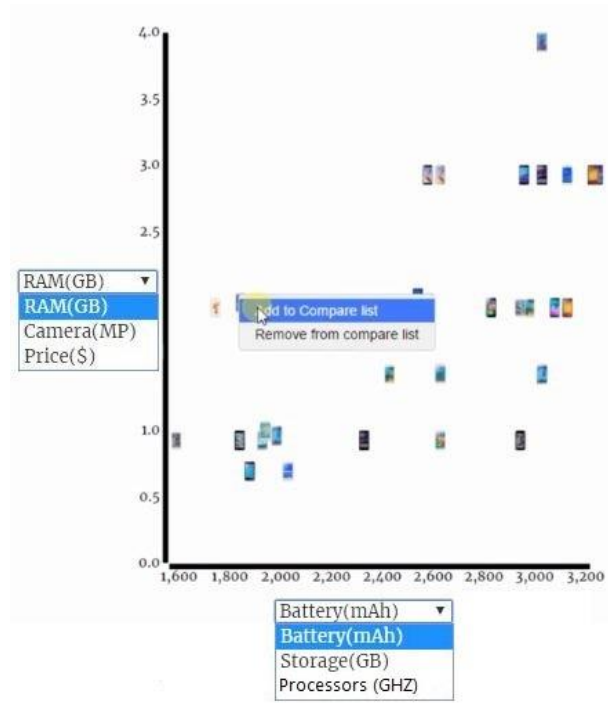

*Figure 5: Scatter plot RAM (Y-axis) and Battery (X-axis)*

Comparative view of multi-attribute products based on their desired attributes using Scatter plot visualization. The user can change the attribute by selecting the desired attribute in the dropdown menu as shown. The purpose of this interface is to allow the user to select a subset of the products for further consideration. It is important to note that 'comparative view' visualization is for comparing ordinal data and will not work on non-ordinal data such as brands. This does not limit its application because if the brand is important to the user then he/she will already select a brand in 'filtering' step, then this visualization will have comparison of ordinal data for the selected brand. On the other hand, if brand is not important then 'comparative view' will have comparison of ordinal data for all brands.

Moreover, the location of the desired product in this view depends on its attributes. For example, if battery (

Figure *5* x-axis) and RAM (

Figure *5* y-axis) are the desired attribute then the region that holds preferred product would be top right of scatter plot but this may change for different attribute such as 'price.

3. **Product View**

The product view model deals with showcasing the model in 3D display along with its distinguishing characteristics. Whenever a user will click on a data item in the scatter plot, its specifications will be loaded. Its objective is to provide a quick detailed description of the product attributes for further consideration of an item. The user can click on the cross button icon shown in Figure 6 at the top right corner to cancel the product view.

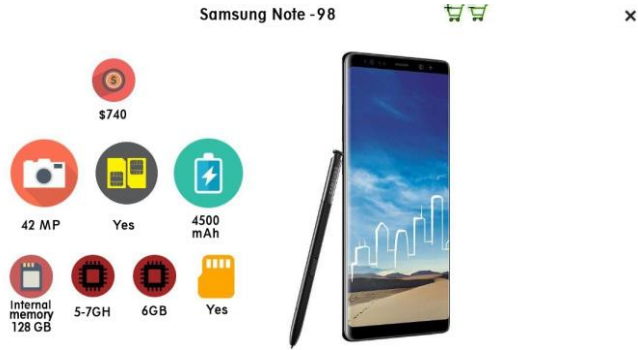

*Figure 6: Product view of a multi-attribute product for detailed consideration.*

This view allows the user to inspect a 3D model of the related product and consider the various attributes in detail. Note that this step gets its icons from a database created by the seller. Icons for common products like laptops and mobile phones are generally understood by users. For more specialized equipment like IoT devices for example, the customers will likely be familiar with the related iconography.

4. *Comparative Visualization*

In the previous steps we compared various products attributes and then shortlisted certain products. In this step, the user will compare the selected attributes of the short-listed products.

In the visualization of this step, every selected attribute is assigned a rectangle for each of its specifications such that the larger the rectangle, the greater the value of that attribute, and vice versa. This visualization helps in fast analysis of various attributes in a single glance. These rectangles are color coded so that it is easier to differentiate between each product. For example in Figure 7, the camera attribute of three different smartphones is shown graphically by three different colors from which we can see that orange color is at the top of the graph window showing that this product is outperforming other products in the comparison graph. At this step, the user makes the final decision and selects the product he/she desires.

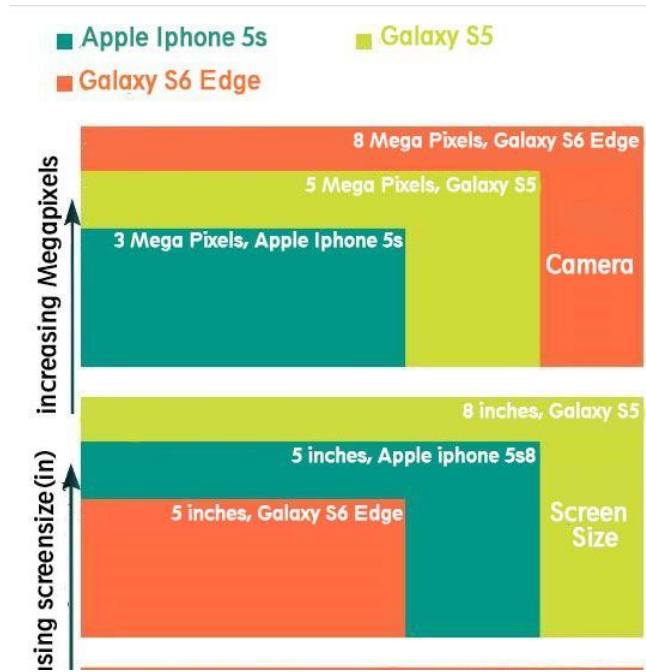

*Figure 7: Comparative visualization of shortlisted products.*

Shortlisted products are color coded and are compared on attributes of interest where each selected attribute is assigned a rectangle. These rectangles are drawn for each product in their respective colors, where biggest rectangle means that an attribute is better in a certain product. For example, in Figure 7 orange rectangles are associated with 'Galaxy S6 Edge'. For attributes where orange is the biggest rectangle it is better than the other shortlisted products. In above example, for attribute 'Camera' 'Galaxy S6 Edge' is the best in comparison to the other two smartphones. Moreover, arrows on the left-hand side indicate the ascending order of the color graphs. This visualization helps in fast analysis of various attributes in a single glance. Like the comparative view, this visualization works with both ordinal and interval data.

# Evaluation

In order to evaluate our proposed visualization pipeline, we conducted a controlled experiment. The discussion in this section is further divided into three subsections: experimental design, threats to validity, and results of the evaluation.

## A. Experimental Design

### 1) Method

To evaluate our proposed visualization pipeline, we first built a prototype of existing version of an e-commerce website shown in Figure 8 containing dataset of hundred smartphones. This typical e-commerce interface allows users to apply various attribute filters to narrow down the search results and then to compare a limited number of smartphones through a *product comparison table* as discussed above and shown in Figure 9. The reason for this choice is twofold. Firstly, Dimara *et al*. [9] empirically showed that tabular representation works best among other possible representation. Secondly, product comparison tables are commonly used in commercial e-commerce portals. Our default interface prototypes allows user to filter products on their attributes based to common GUI components.

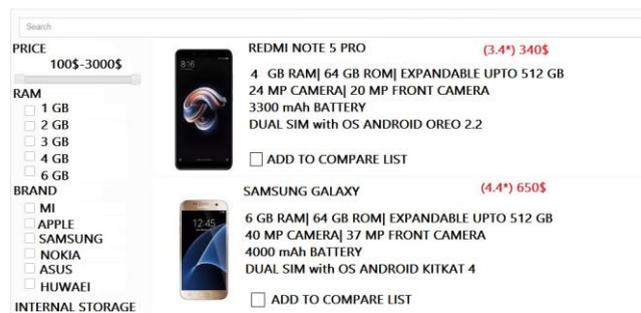

*Figure 8: Typical interface of an E-commerce website.*

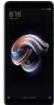

*Figure 9: Product comparison table in a typical ecommerce websites.*

Next, using the same multi parametric product i.e. smartphone, we used a different dataset of hundred smartphones while incorporating our visualization pipeline in the second e-commerce website prototype thus allowing users to make a smartphone selection using visualization pipeline discussed in Section III. A participant might learn the correct answer during one prototype interface which can influence his/her performance on the second interface prototype on the same task. To cater for such a possibility, we added different phone data in the database of both the prototypes.

*2) Hypothesis Formulation*

Following hypothesis are formulated.

$$H_11: \quad H_0: \mu = \mu'$$

$$H_01: \quad H_1: \mu > \mu'$$

We let μ denote the true mean time *(sec)* of users on the typical e-commerce interface and let μ' denote the true mean time *(sec)* on the proposed visualization interface.

$$H_12: \quad H_0: \mu = \mu'$$

$$H_02: \quad H_1: \mu < \mu'$$

We let μ denote the true mean number of correct responses of users on the typical e-commerce interface and let μ' denote the true mean on the proposed visualization interface.

*3) Participants*

We evaluated the usability of shopping portals for multi-attribute products, by conducting a controlled experiment with thirty participants on our two designed e-commerce prototypes. The summary of demographic information is given in Table 1. We also provided computer efficacy questionnaire [62] to each participant of our study, the details of computer efficacy of the participants is in Appendix A.

*Table 1: Demographic table*

| Category | Values |
|---|---|
| No. of participants | 30 *(19M, 11F)* |
| Age group | All between 18 to 42 years |
| Level of education | All high school graduates |
| Online shopping experience | All having prior experience |

At the beginning of the experiment, we informed participants that they would be required to search specific smartphones having certain criteria and complete the assigned tasks. Also, a brief introduction about both the prototypes was given to the participants of our study. We note here from our literature review that [20] categorized users as either satisfiers or maximizers. Here, the maximizers want to optimize certain qualities of a product and extensively search for such a product [31][29][30]. For example, a user might give more importance to the camera resolution, screen size, RAM and dual SIM options in a smartphone and not be concerned with the operating system or LTE options. Similarly, some company might want to consider laptops from a set of certain manufacturers and with a specific graphics card etc. Since [20] reported that maximizers take the most time to finalizing their purchase, we decided that our control experiment should closely mimic their shopping experience. This was done by designing our tasks such that the user is asked to select a product through some form of multi-attribute optimization, e.g. asking the user to find a mobile phone in a certain price range and highest memory size.

We note here that a generic filter function was available in the typical e-commerce interface but not in the visualization-based interface. The advantage of the visualization interface was not due to a filter function but its ability to allow human to understand the data quicker and better.

*4) Tasks*

In order to make them familiar with the interfaces we asked them to perform 2 trial tasks which were as follows:

- Find an iPhone with the lowest price
- Find the highest priced Samsung phone with 6GB RAM

The tasks that were given to the users were designed to be of two types: simple and complex. In a *simple task,* a user was asked to satisfy constraints on 2 or fewer attributes, while in a *complex task* a user was asked to satisfy 3 or more attributes. For example, finding a smartphone within a certain budget is considered a simple task as it considers only one variable, while if a user is trying to satisfy constraints on camera resolution, RAM and battery capacity then we classify it as a complex task. More specifically, each participant of our study was asked to perform the following tasks.

- *Simple Task 01*: Find a smartphone having battery capacity greater than 2000mAh
- *Complex Task 01*: Find the lowest priced Android smartphone with at least 4GB RAM and camera above 20Mp
- *Complex Task 02*: Find Samsung smartphone of type either Note or Edge that has the highest MP camera
- *Complex Task 03*: Find a smartphone that has: RAM greater than 3 GB, camera greater than 16MP, and battery more than 3000mAh

*5) Evaluation Metrics*

We adopted the method similar to [21][22] for evaluating our proposed visualization. It compares usability of the proposed interface with an existing e-commerce interface. The number of participants that are selected are usually small as it has been shown that a small set participant uncover most of the usability problems in an experimental setting. Usability is assessed by measuring the following three components of usability [23][24]:

- *Efficiency:* time required by a user in performing the given tasks, measured in seconds
- *Effectiveness:* by noting the number of correct answers by participants while performing tasks on both the interfaces i.e. typical e-commerce interface and proposed visualization interface
- *Satisfaction:* we adapted the IBM standard survey [63] to collect participants subjective views on the proposed visualization pipeline i.e. proposed interface

*6) Experiment Session*

After understanding the instructions at the beginning, each participant started their task by filling some basic information on the home page. There was no time limit for participants. They were also asked not to use their mobile phones during the experiment to avoid any disturbance or distraction.

We conducted this study with participants in the month of December 2018. Timing was adjusted according to the participant's convenience; however, they were asked to choose a time between 9 am to 5 pm of the reserved day. We performed our experiment in a computer lab and also provided all our participants the same laptop of 15.5" screen with a resolution of 1366x768 pixels. Also, all other applications were closed on the testing laptops to ensure that computer performance was not affected by other applications.

B. *Threats to Validity*

This subsection of the paper has addressed the threats to validity and the way we attempted to mollify them during the experiment conducted. Following internal validity threats have been addressed in our study: **Ordering effect:** We countered the effect of ordering; half of the participants of our study first performed the given tasks on the typical e-commerce interface and remaining started from the proposed interface. **Uncontrolled variation**: We conducted the controlled experiment during the same time of day and in the same location *(a quite lab)* to avoid uncontrolled variation. Moreover, all participants of our study were Windows computer users and all of them performed the tasks on the same windows-based laptop. This ensured that any uncontrolled variation due to software or hardware is minimized. **Selection bias**: Selection of participants for the experiment was based on convenience sampling including both male and female with different age groups. Hence, the results of this study cannot be generalized to the whole population.

The threats to external validity related to our experiment are *training validity* and *sample not representative of the population*. **Training validity**: We assured this threat to external validity by demonstrating participants on how to begin with their tasks involved in the experiment as well as by letting them to perform trial tasks before proceeding to the actual tasks. We also informed them to ask any question at any stage during our experiment. **Sample not representatives of the population**: The participants contributed in this study, were truly the representatives of the population. All our participants had previous e-shopping experience thus negated the presence of this external validity threat. Next, we highlighted threats to conclusion together with construct validity. We selected a sample of 30 for the experiment in this research. Henceforth, the sample size of our study results as a threat to conclusion validity. Furthermore, "*Construct validity is the degree to which an experiment correctly captures the intended measurements*" [64]. The objective of this study is to measure the usability of two interfaces for e-commerce websites in the perspective of end users. Since usability is considered a subjective concept, it is generally agreed that efficiency, effectiveness, and satisfaction are the prime independent constructs of usability. Therefore, in our research study, we decided to measure usability in terms of those parameters.

## Results and Analysis

To draw meaningful conclusions from our research experiment, we applied statistical analysis method on the gathered data, shown in Table 2 to interpret the results. Furthermore, we have explicitly discussed what these three usability constructs i.e. *efficiency*, *effectiveness* and *satisfaction* mean in our context and how we measured them during the experiment.

*Table 2: Data gathered from the controlled experiment*

| Tasks | Participants average time *(sec)* on typical e-commerce interface | Participants average time *(sec)* on proposed visualization interface | No. of successful participants on typical e-commerce interface | No. of successful participants on proposed visualization interface |
|---|---|---|---|---|
| **ST01** | 52 sec | 51 sec | 28 | 28 |
| **CT01** | 73 sec | 59 sec | 15 | 18 |
| **CT02** | 94 sec | 53 sec | 18 | 22 |
| **CT03** | 64 sec | 43 sec | 21 | 27 |

*A. Efficiency*

We first calculated the 95% confidence interval for all four tasks. Figure 10 shows these results which clearly shows that the confidence intervals overlap in case of the simple tasks. While for the complex tasks they don't overlap hence it is possible that they are statistically significant. To see how our proposed interface performs overall, we calculated the mean time taken by our participants on the typical e-commerce interface and the mean taken on the proposed interface. The mean time taken by our participants on the typical e-commerce interface was 283.43 sec (s.d. 90.31) and the mean time taken by our participants on the proposed interface was 205.83 sec (s.d. 85.83). This mean time was calculated for all user tasks discussed above. Figure 11 shows the results along with 95% confidence intervals. Since the confidence intervals do not overlap hence, we apply t-test to ensure the statistical significance of these results.

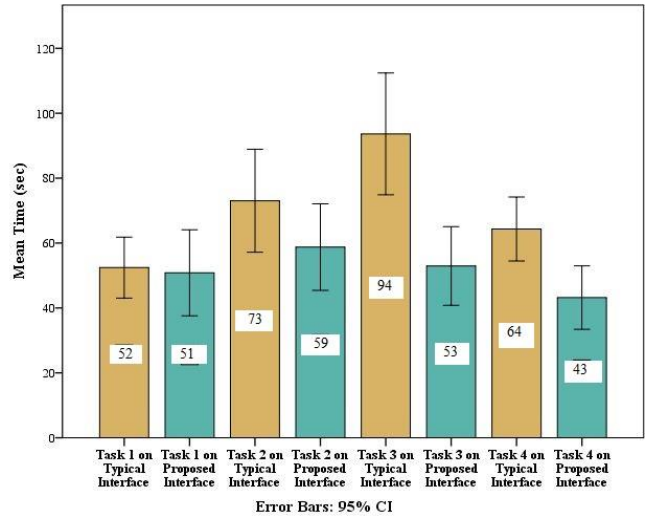

*Figure 10:* **Task-wise efficiency results. We can see that the confidence intervals overlap in case of the simple tasks. While for the complex tasks they don't overlap hence it is possible that they are statistically significant.**

We concluded that the mean time of participants on the proposed interface is less than the mean time of participants on typical e-commerce interface as our t-test gave us a p-value of 0.002 at 5% level of significance ($\alpha=0.05$). Hence, we adopt the alternative hypothesis.

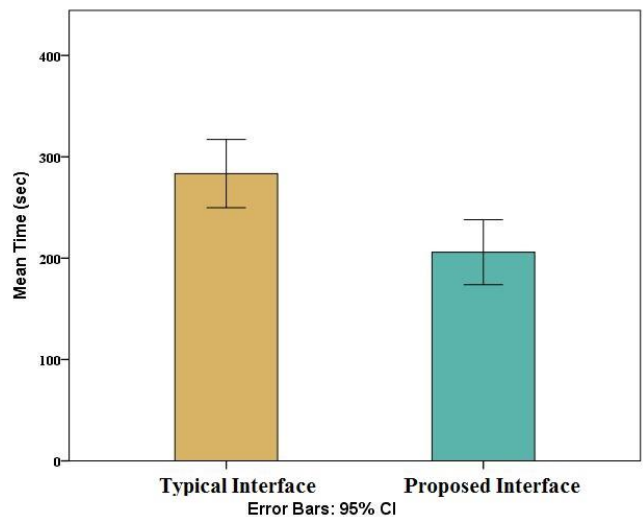

*Figure 11:* **Meantime (sec) of participants on both the interfaces. Typical e-commerce interface meantime is 283.43sec while proposed interface meantime is 205.83 sec.**

## B. Effectiveness

We first noted the number of correct responses by each participant for the 4 tasks on the typical e-commerce interface as well as on the proposed interface. We then calculated the mean number of correct responses in each case. Figure 12 shows the results along with 95% confidence intervals. However, the confidence intervals overlap as shown in the figure. Hence, we can conclude that statistically there is no difference between the effectiveness of both interfaces. We note here that the mean accuracy of the typical e-commerce interface is 68.33%, and the mean accuracy of the proposed interface is 78.30%. So, although, not statistically more effective, the participants of our study, on average, still selected phones that met the required criteria in 10% more tasks. This better performance can be more clearly observed from Figure 13.

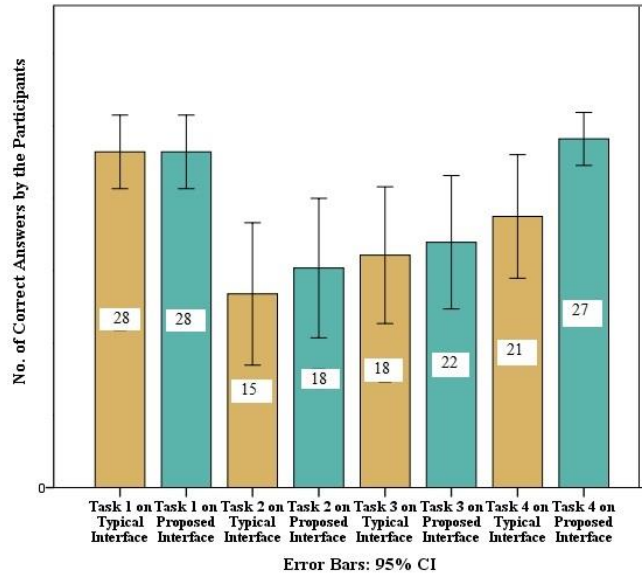

*Figure 12: Number of correct answers by the participants on both the interfaces on each task. We can see that the confidence intervals overlap hence statistically speaking they are equally effective.*

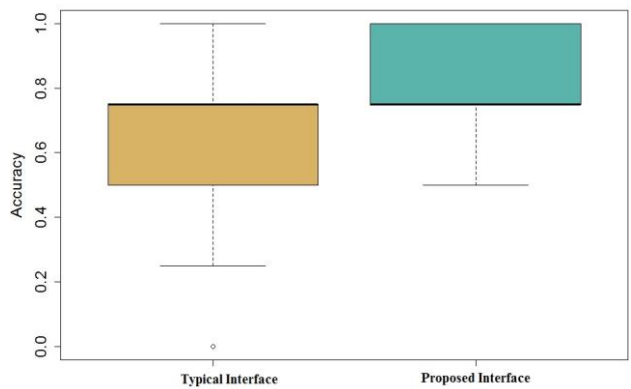

*Figure 13: Accuracy level of correct answers by the participants on both the interfaces. The mean accuracy of the typical e-commerce interface is 68.33%, and the mean accuracy of the proposed interface is 78.30%.*

### C. Satisfaction

We adapted the IBM standard survey [63] to collect participants subjective views on the proposed visualization interface to evaluate our third usability parameter of user satisfaction. We used two points Likert scale i.e. Agree or Disagree to measure participant's satisfaction towards the new e-shopping interface. The results of which are summarized in Figure 14. We concluded that almost all our participants had a satisfactory opinion to the proposed visualization interface. For example, first question was "I feel comfortable using the new interface". 77% were agree with this question while 23% were disagree on it.

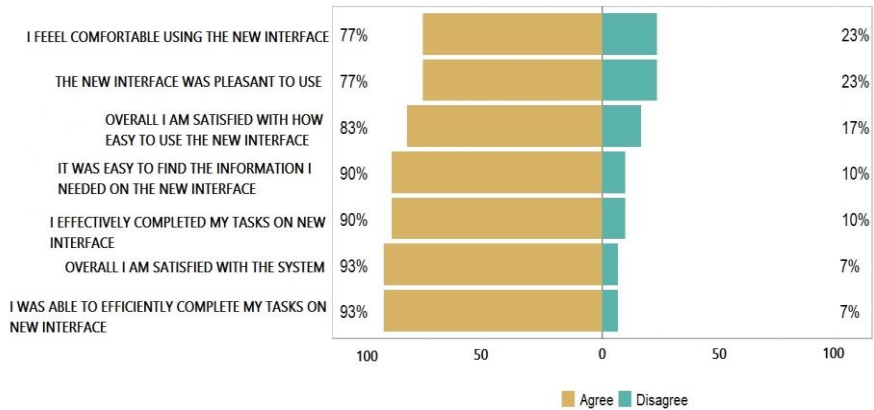

*Figure 14: Satisfaction questions and summary of participant's responses. We can clearly see that most participants were satisfied with the proposed interface.*

## Conclusion

Information visualization is a way of representing some form of collected data. The core objective of which is to extend cognition and perception of a human mind as it helps to analyze and reason about data in a more efficient and effective manner. However, there is less adoption of visualization concept on e-shopping portals while depicting products data to its users. Because of which users find it difficult to shortlist and compare multi-attribute products for final decision-making. Therefore, in our research, we proposed a new visualization-based interface for e-commerce websites that makes the task of finding and comparing multi-attribute product/service easier for the user thus makes the overall process of product purchase decision-making time-efficient and effective. Our proposed visualization pipeline aimed to address the following usability parameters *efficiency*, *effectiveness,* and *satisfaction*. Later in our research, we conducted a controlled experiment to evaluate our solution for e-commerce websites. The results of which are shown in Table 2.

The restriction implies to this work corresponds to the validation of our proposed solution. We evaluated our proposed visualization interface on a sample size of 30 selected participants based on convenience sampling, which results as a threat to conclusion validity. Hence, the results of this study cannot be generalized to the whole population. However, the results of our experiment showed that the visualization-based interface results in reducing the time taken to decide products to buy while maintaining the effectiveness of the typical e-commerce interface. Similarly, the satisfaction survey showed that users generally preferred the visualization-based interface. Hence with further research and improvement, it is possible that similar interfaces might one day be deployed in e-commerce websites.

Therefore, there is a need to conduct detailed user studies to deduce generalizable results. In future, we would work to make our results more generalized by evaluating it on a bigger sample. Furthermore, our visualization is specifically designed for e-shopping websites and would not be easily adapted to other websites. In future, another research work would be to redesign our visualization method so that it can be used on websites of other domains as well.

## Acknowledgement

The proposed visualization interface was initially implemented as a Final Year Project by Maleeha Afzal, Syeda Rija Zainab and Zahra Shaukat under the supervision of Muhammad Muddassir Malik of NUST School of Electrical Engineering and Computer Science, Islamabad. Furthermore, in this research the idea was re-implemented as an e-commerce website prototype by Shoaib Masood of COMSATS University, Islamabad, Pakistan.

Based Decision Support in Lighting Design," *IEEE Trans. Vis. Comput. Graph.*, vol. 22, no. 1, pp. 290–299, 2016.

[56] C.Williamson and B. Shneiderman., "The Dynamic HomeFinder : Evaluating Dynamic Queries in a Real-Estate Information Exploration System," in *In Proceedings of the 15th Annual International ACM SIGIR Conference on Research and Development in Information Retrieval, SIGIR*, 1992, pp. 338–346.

[57] C. A. and B. Shneiderman, "Visual Information Seeking : Tight Coupling of Dynamic Query Filters with Starfield Displays," in *In Proceedings of the SIGCHI conference on Human factors in computing systems*, 1994, pp. 313–317.

[58] A. Sarikaya, S. Member, and M. Gleicher, "Scatterplots : Tasks , Data , and Designs," vol. 24, no. 1, pp. 402–412, 2018.

[59] G. Carenini and L. John, "ValueCharts : Analyzing Linear Models Expressing Preferences and Evaluations," in *In Proceedings of the working conference on Advanced visual interfaces*, 2004, pp. 150–157.

[60] S. Gratzl, A. Lex, N. Gehlenborg, H. Pfister, and M. Streit, "LineUp : Visual Analysis of Multi-Attribute Rankings," *IEEE Trans. Vis. Comput. Graph.*, vol. 19, no. 22, pp. 2277–2286, 2013.

[61] R. Netzel, J. Vuong, U. Engelke, S. O. Donoghue, D. Weiskopf, and J. Heinrich, "Visual Informatics Comparative eye-tracking evaluation of scatterplots and parallel coordinates," *Vis. Informatics*, vol. 1, no. 2, pp. 118–131, 2017.

[62] O. Khorrami-Arani, "Researching computer self-efficacy," *Int. Educ. J.*, vol. 2, no. 4, pp. 17–25, 2001.

[63] J. Lewis, "IBM computer usability satisfaction questionnaires: Psychometric evaluation and instructions for use," *Int. J. Hum. Comput. Interact.*, vol. 7, no. 1, pp. 57–78, 1995.

[64] N. Ahmad, A. Rextin, and U. E. Kulsoom, "Perspectives on usability guidelines for smartphone applications: An empirical investigation and systematic literature review," *Inf. Softw. Technol.*, vol. 94, no. September 2017, pp. 130–149, 2017.
**Appendix A**

The responses of which were gathered on a three-point Likert scale i.e. (Agree, Neutral, and Disagree). The collective results of which are summarized in

Figure15. We later calculated the net computer efficacy score of each participant by replacing Agree with a numeric value 1, Neutral with 0 and Disagree with -1. The overall net computer efficacy score had a mean of 0.55 and a standard deviation of 0.01.

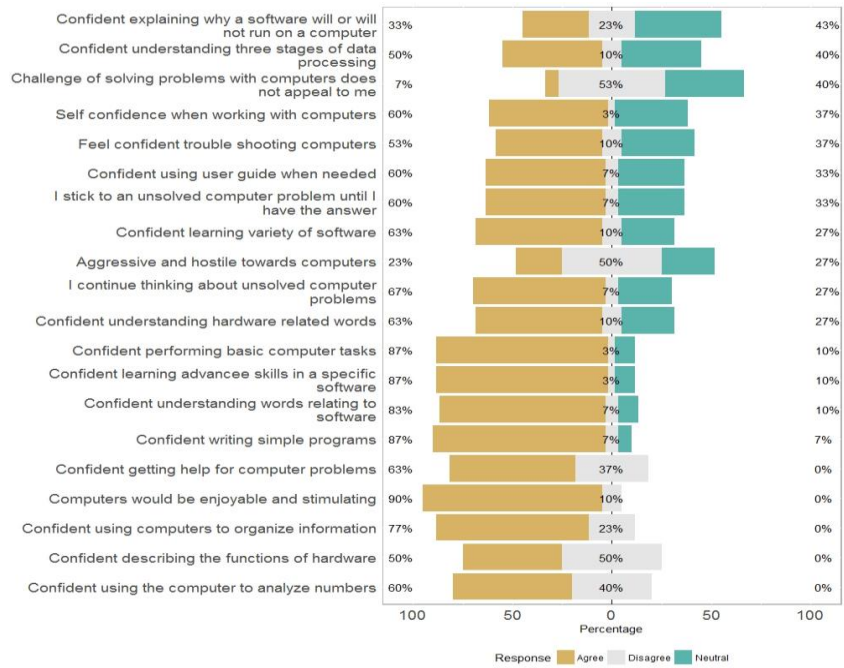

*Figure15:* **Computer efficacy questions and summary of participant's responses. Users in general are comfortable in using computers.**